\documentclass[prl,twocolumn,aps]{revtex4}

\usepackage{amssymb}
\begin{document}
\bibliographystyle{revtex}

\title{Orbital effect of a magnetic field on the low temperature state in the
       organic metal $\alpha$-(BEDT-TTF)$_2$KHg(SCN)$_4$}
\author{D.~Andres$^1$}
\author{M. V.~Kartsovnik$^1$}
\author{W.~Biberacher$^1$}
\author{H.~Weiss$^2$}
\author{E.~Balthes$^2$}
\author{H.~M\"uller$^3$}
\author{N.~Kushch$^4$}

\affiliation{$^1$Walther-Meissner-Institut, Bayerische Akademie
der Wissenschaften, D-85748 Garching, Germany}

\affiliation{$^2$ High Magnetic Field Laboratory, MPI-FKF and CNRS,
BP 166, F-38042 Grenoble Cedex 9, France}

\affiliation{$^3$European Synchrotron Radiation Facility, F-38043
Grenoble, France}

\affiliation{$^4$Institute of Problems of Chemical Physics,
Ru.Ac.Sci., 142432 Chernogolovka, Russia}

\begin{abstract}

The effect of pressure on the {\it B--T} phase diagram of
$\alpha$-(BEDT-TTF)$_2$KHg(SCN)$_4$ is studied. The measured phase lines
can be well described by a recent model of a charge-density wave system
with varying nesting conditions. A remarkable increase of the transition
temperature with magnetic field is found in a certain pressure and field
range. We associate this result with a dramatic enhancement of the
orbital effect of magnetic field due to a deterioration of the
nesting conditions by pressure. Furthermore, we present data which
can be interpreted as a first sign of field-induced charge-density
waves.

\end{abstract}
\maketitle

The organic charge transfer salt
$\alpha$-(BEDT-TTF)$_2$KHg(SCN)$_4$ has a layered crystal structure,
consisting of conducting bis(ethylenedithio)tetrathiafulvalene and
insulating anion sheets \cite{ish98}, that leads to a strong anisotropy
of the electronic system.
Numerous anomalies displayed by this material in magnetic
field can reasonably be explained by a density-wave instability
of the quasi-one-dimensional (Q1D) part of the electronic system and
its interaction with the quasi-two-dimensional (Q2D) conducting band
(see for a review \cite{wos96}).

Extensive studies of the magnetic field-temperature ({\it B--T\/})
phase diagram \cite{kar97,kov98,bis98,chr00,qualls00} have
provided a substantial argument for the charge-density-wave
(CDW) nature of the low-temperature state in this compound.
Both thermodynamic \cite{kar97,kov98,chr00}
and interlayer transport \cite{kar97,bis98,qualls00}
measurements show a decreasing transition temperature $T_p$
with increasing field. Such a behavior
is expected for a well nested CDW system \cite{ken97,bie99}.
This is due to the competition between the Pauli
paramagnetism and the CDW instability in a magnetic
field. For a perfectly nested CDW this Pauli effect causes a
gradual decrease of the transition temperature with field. In the
low field limit ($B\! <\! <\! B_c \sim [k_BT_p(B=0)] /\mu_B$),
$\Delta T_p / T_p$ is proportional to $B^2$ \cite{die73}. At
higher fields, $B\approx B_c$, when the Zeeman energy reaches the
value of the zero temperature energy gap, theory proposes a first
order phase transition at low
temperatures: the perfectly nested CDW state transforms into a
CDW/SDW hybrid state with a shifted, field dependent
nesting vector \cite{ken97,bie99}. In the present compound the
transition temperature is remarkably lower than in most known CDW
systems. This gives the unique opportunity to extend
the studies of field effects far beyond the low field limit
even in static magnetic fields. Indeed, previous
experiments \cite{kar97,bis98,chr00,qualls00,sas96} have
already demonstrated the existence of a new phase at fields above 24~T and
temperatures below 4 K, which can be associated with the CDW/SDW state.

Another, so-called orbital effect of magnetic field must be taken
into account in an imperfectly nested density-wave system.
Under a magnetic field applied parallel to the open sheets of the Fermi
surface, the electrons are
forced to move in $k$-space perpendicular to the field along the
sheets, causing an oscillatory motion in real space which becomes
more restricted to one dimension with increasing the
field. This should cause a stabilization of the density-wave
state and lead to an increase of $T_p$ with field
\cite{mon88}. Such an increase due to the orbital effect has been
observed in SDW systems \cite{kwa86}. Moreover, the
orbital quantization was shown to lead to a
fascinating macroscopic quantum phenomenon known as
field-induced spin-density waves (FISDW)
(see for a review \cite{cha96}). Despite
similar predictions \cite{bie99}, there exists up to now no clear
evidence of the orbital effect on $ T_p$ in a CDW system
\cite{comment2}. In $\alpha$-(BEDT-TTF)$_2$KHg(SCN)$_4$, the phase diagram
can be fairly well described by a dominant Pauli effect of magnetic field
on a CDW state, although a weak dependence of $T_p$ on the magnetic field
orientation \cite{kov98,chr00} may be interpreted as an indication
of a small orbital effect. The theoretically predicted \cite{bie99}
competition between the two effects was recently suggested
by Qualls et al. \cite{qualls00} to be
a reason for a significant modification of the phase diagram at
magnetic fields strongly tilted towards the conducting layers.
However, other studies of the present material at high tilt angles
\cite{icsm} reveal a complicated behavior which does not fit
into the simple picture proposed in Ref. \cite{qualls00}.

In this work we report on a direct manifestation of the orbital
effect on the CDW phase diagram obtained by tuning the
nesting conditions in $\alpha$-(BEDT-TTF)$_2$KHg(SCN)$_4$
by quasi-hydrostatic pressure.
The interplane resistance of $\alpha$-(BEDT-TTF)$_2$KHg(SCN)$_4$ was
measured at temperatures down to 0.4~K in magnetic fields up to 28~T,
directed perpendicular to the layers, at different pressures up to
$P =$~4.6~kbar. Quasi-hydrostatic pressure was applied
using either a conventional clamp cell or a He-pressure
apparatus. Several samples from different
batches were measured, revealing basically the same
behavior.

Although the determination of transition points from the
magnetoresistance is not straightforward, reasonable
estimates in an applied magnetic field
can be made via Kohler's rule, which is a similarity
law for the magnetoresistance \cite{pip89}. It is based on the
assumption that the scattering processes do not depend on
magnetic field. If one further expects the zero field resistance
$R_0$ to be inversely proportional to the scattering time $\tau$,
the magnetoresistance can be expressed as a general
function of $B\! /\! R_0$:
\begin{equation}
[R_B(T)-R_0(T)]/R_0(T)= F[B/R_0(T)],
\end{equation}
that constitutes Kohler's rule. Kohler's rule has
already been found to work well in several organic
metals (see e.g. \cite{ham87}). Fig.~1 shows typical
scaling plots, so-called Kohler plots, obtained from
temperature sweeps at fixed magnetic
fields which are depicted in the inset. It is clearly seen that
at higher temperatures the curves follow one general function, in
accordance with Kohler's rule. This suggests that this
rule is valid for the NM state of $\alpha$-(BEDT-TTF)$_2$KHg(SCN)$_4$ at
the given orientation and range of the applied field. At
lower temperatures all the curves start to diverge dramatically.
This is consistent with an earlier report on a strong violation
of Kohler's rule in the LT state of this material \cite{ken98}.
We therefore ascribe the deviation from Kohler's rule to the
phase transition from the NM to the LT state. As a
characteristic  temperature of the transition $T_p$, we take the
temperature corresponding to the crossing point of linear
extrapolations from the NM and LT parts of the Kohler plots as
shown in Fig.~1 for $B = 10$~T. The dependence of $T_p$ on
magnetic field is shown in Fig.~2 (empty circles) for three different
pressures. We note that the same behavior is obtained for the temperatures
corresponding to the maximum curvature of the Kohler plots or a
typical kink in their derivatives. Thus, even though we cannot
assert an exact definition of the absolute value of the critical
temperature from the above procedure, we believe
that the curves in Fig.~2 reflect the correct dependence of the
real critical temperature on magnetic field and hydrostatic
pressure.

At ambient pressure, the observed monotonic shift of
the transition temperature to lower values with increasing field
is consistent with data obtained by specific heat \cite{kov98}
(squares in Fig.2) and magnetic torque experiments \cite{chr00}.
Under pressure the phase boundary moves to lower temperature; no
transition has been detected at 4.6 kbar in agreement with
previous studies \cite{han96}. Furthermore,
pressure causes a remarkable change in the shape of the phase
boundary: in a certain range the field clearly \emph{stabilizes}
the LT state. This result can be readily understood in terms of a
competition between the orbital and Pauli effects of
magnetic field on the CDW state. At
ambient pressure, when the nesting is good, the Pauli
paramagnetic effect dominates, leading to a constant decrease of
$T_p$ with increasing field. An applied pressure
deteriorates the nesting conditions,
thereby suppressing the zero-field $T_p$. At the same
time, the orbital motion in a magnetic field
perpendicular to the ac-plane acts to
effectively reduce the dimensionality
of the electronic system,
resulting in a relative increase of $T_p$. Thus, with an
increasing pressure, hence warping of the Fermi surface,
the orbital effect becomes more pronounced and may even
become dominant as seen in Fig.~2. However, if the characteristic
frequency of the orbital motion, $\omega _c =eB\nu
_F^{(1D)}d/\hbar c$ ($\nu _F$, Fermi velocity; $d$,
length of the unit cell along the open sheets of the Fermi
surface; $c$, velocity of light), is sufficiently
high, $\hbar \omega _c
> k_BT_p(0)$ , the contribution from the orbital effect to
$T_p$($B$) saturates \cite{sas96} and the CDW state should be
eventually suppressed due to the Pauli effect.
This qualitative consideration is found to be in very good
agreement with the evolution of the {\it{B--T}} diagram of
$\alpha$-(BEDT-TTF)$_2$KHg(SCN)$_4$ under pressure.

In Fig.~3 we present the phase lines obtained from the
investigation of several samples under different pressures in
magnetic field up to 27~T. The data points correspond to one
single experiment for each pressure. The circles are taken from
the Kohler plots as described above. The triangles represent the
so-called kink transition (that is supposed to be a transition from the
low-field CDW to the high-field CDW/SDW hybrid state
\cite{bis98,chr00,qualls00,ken97}),
which was recorded in the field
sweeps of the magnetoresistance at $P$~=~0 and 1.8~kbar. The solid
line at ambient pressure illustrates
the general behavior observed in previous works
\cite{kar97,kov98,bis98,chr00}. As a whole, the phase diagrams
are strikingly similar to those predicted by Zanchi et al.
\cite{bie99} for a CDW system with varying nesting conditions.
The latter are shown in the inset in Fig. 3. Here, the
imperfect nesting is introduced by the second order transfer
integral $t_c'$ entering the dispersion relation:
\begin{equation}
\epsilon=\displaystyle \nu_F(|k_x|-k_F)-2t_c
\cos{(k_cc)}+2t_c'\cos{(2k_cc)},
 \label{eldis}
\end{equation}
and $t_c'^*$ is a critical value of $t_c'$ at which the CDW
is completely suppressed at zero field ($t_c'^*$ can be
estimated as $\cong k_BT_p^0(0)$ where $T_p^0(0)$ is the
zero-field transition temperature at $t_c'=0$ \cite{has86}).
The transition between the low-field CDW and high-field
  hybrid CDW/SDW states was analysed so far only for a perfectly
  nested system ($t_c'$=0) \cite{bie99}. Therefore the phase lines
  in the inset in Fig. 3 do not include this transition. It would be
  highly interesting to extend these studies for the case of finite
  $t_c'$ and to compare the theory with the pressure dependence of
 the experimentally observed kink transition.

An explicit comparison between the experiment and theory cannot
be done at this stage: On the one hand, the boundaries can change
to some extent depending on the values of the coupling constants.
On the other hand, the theoretical model \cite{bie99} ignores
such factors as fluctuations and the presence of the additional,
Q2D conducting band which can also lead to a modification of the
phase diagram. Nonetheless, certain conclusions based on the
qualitative similarity between the experiment and theoretical
predictions shown in Fig.~3 can be made.
At $P=3.6$~kbar the NM state persists down to at least 1.4~K at
fields below 10~T, whereas clear deviations from Kohler's rule are
detected at $B>12$~T. This behavior obviously corresponds to
$t_c'/t_c'^* >1$.
The competition between the orbital and Pauli effects is expected to
be the most pronounced at $t_c'/t_c'^*= 1.0\pm 0.1$
\cite{bie99}. It is this region in which both $T_p(0)$ and the
shape of the phase line are extremely sensitive to $t_c'$. Our
data in Fig.~3 suggest that the pressure of 2.3~kbar corresponds to
$t_c'/t_c'^*$ almost exactly equal to 1.

The non-monotonic field dependence of $T_p$ at 2.3~kbar is also
reflected in isothermal field sweeps of the magnetoresistance
which are shown in Fig.~4. In contrast to a smooth behavior at
$T\geq 5$~K, the magnetoresistance at $T=3.6$~K exhibits a clear
enhancement due to entering the LT state. According to the phase
diagram in Fig.~3, the LT state occurs at 3.6~K in the field
range between $\approx$~6.5 and 16.5~T as indicated in Fig.~4 by
dashed lines. The low-field feature
rapidly weakens and shifts to lower fields (as marked by the
dotted vertical arrows in Fig.~4) as the temperature is
reduced below 3~K \cite{comment}.
The decrease of the magnetoresistance background at
high fields manifests a field-induced transition into the
high-field modification of the LT state (kink transition) or into
the NM state and can be observed in the field sweeps at any
temperature below 5~K.

At $T<$~2.5~K a hysteresis in the magnetic field sweeps
emerges in a broad interval as marked by solid
vertical arrows in Fig.~4. It is accompanied by a change
of the slope of the magnetoresistance in a certain field
range as clearly
seen at the 0.5~K curve at 4-5~T. Both the hysteresis and
non-monotonic behavior of the magnetoresistance become even more
pronounced at higher pressure as shown in Fig. 5 for $P=3$~kbar,
$T=1.4$~K.
These anomalies, at first glance surprising, may turn
out to be a sign of an interesting quantum phenomenon. For
certain nesting conditions, namely when $t_c'$ becomes comparable
to $t_c'^*$, theory \cite{bie99} predicts a cascade of
field-induced CDW (FICDW) transitions. This phenomenon is
analogous to already well known FISDW \cite{cha96}, where the
quantized adjustment of the nesting vector serves to keep the
Fermi energy level between the Landau levels, in order to
stabilize the SDW in a varying magnetic field.
Comparing our results with the theoretical prediction \cite{bie99},
we suggest that the features displayed in Fig. 4 (for the lowest
temperatures) and Fig. 5 may be a manifestation of the
FICDW phenomenon. In this context, the magnetoresistance
behavior at 3~kbar (Fig. 5) can be described as follows: The increase
of the slope at 2.7 T corresponds to the boundary between the NM and
FICDW regions. Due to the relatively high temperature \cite{leb00},
no clear features can be resolved between 3 and 5 T. The enhancement
of the magnetoresistance at $\approx 5$ T is the first
direct indication of switching between different FICDW subphases. As
the field further increases, the transition anomaly (at
$\approx 7.4$ T) becomes sharper and exhibits a pronounced
hysteresis. If we associate the transition points with the maxima
in the $d^2R/dB^2$ dependence shown in the inset to Fig. 5, and assume
them to be periodic in $1/B$, another transition at $B\cong 12.5$ T
might be expected. However, no clear anomaly has been found between
10 and 15 T. On the one hand, this may be an indication that the
system is already in the $n=0$ state (i.e., the
longitudinal component of the nesting vector is $q_{\|}=2k_F$
\cite{cha96}) above
7.4 T. On the other hand, the FICDW transition above 10 T should be
influenced by i) the orbital quantization of the Q2D band reflected in
strong Shubnikov-de Haas oscillations; and ii) the Pauli effect
of the magnetic field which is expected to induce the kink transition well
below 20~T at the given pressure. Further theoretical and experimental
studies should clarify how these two mechanisms interfere with the
field-induced quantization of the nesting vector.

Finally, we conclude that although the data shown in
Figs.~4 and 5 are not yet sufficient to claim
unambiguously the detection of FICDW, the similarities between the
experimentally obtained phase diagram and that proposed
theoretically for a CDW system \cite{bie99} make the present
material a promising candidate for the realization of
this new quantum phenomenon.

Summarizing, the {\it B--T} phase diagrams of
 $\alpha$-(BEDT-TTF)$_2$KHg(SCN)$_4$ at different pressures can be
consistently interpreted in terms of the interplay between the Pauli and
orbital effects of the magnetic field on a CDW system with varying nesting
conditions. The orbital effect at pressure of $\simeq 2$ kbar is clearly
manifested by a remarkable increase of the transition temperature
in magnetic field. The non-monotonic hysteretic behavior of
the magnetoresistance at pressures corresponding to
$t_c'/t_c'^*\gtrsim 1$ provides an argument for the existence of
FICDW subphases.

The work was partially supported by the TMR program of EU, grant
ERBFMGECT 950077 and by the DFG-RFBR grant 436 RUS 113/592/0-1 (R).


\newpage

FIG.~1: Kohler plots at: a)~ambient pressure and
b)~1.6~kbar. The corresponding resistance versus
temperature curves at fixed magnetic fields are shown in the
insets.

FIG.~2: Phase boundaries between the LT and NM states
 obtained from the Kohler plots at three different pressures (sample~\#1,
 circles) and from earlier specific heat measurements at ambient
 pressure \cite{kov98} (squares).

FIG.~3: {\it B--T}~phase diagrams measured at 0, 1.8,
3.6~kbar (sample~\#2) and 2.3~kbar (sample~\#3). Circles:
boundary between the NM and the LT states; triangles: kink
transition. Inset: theoretically proposed phase diagrams of a CDW
system at different nesting conditions \cite{bie99}.

FIG.~4: Magnetoresistance of sample~\#3 under $P=2.3$~kbar,
at several temperatures. Vertical dashed lines correspond to the
phase boundaries in Fig.~3. The features marked by arrows
are discussed in the text.

FIG.~5: Magnetoresistance of sample~\#4 at 1.4~K under $P=$~3~kbar.
Inset: The second derivative $d^2R(B)/dB^2$ taken after filtering out the
Shubnikov-de Haas signal, vs. inverse field.


\begin{thebibliography}{10}
\bibitem{comment}
\bibinfo{note}{Noteworthy, although the phase diagram presented in Fig.~3
  suggests $T_p$($B$)~$\gtrsim$~3~K, the low-field feature can be resolved
  in the field sweeps at temperatures down to at least 2~K. Taking into
  account a smooth character of the changes in the magnetoresistance during
  the NM-LT transition, one could suggest that the existence of this feature
  below $T_p$ is due to the proximity to the phase boundary.}
\end{thebibliography}
\end{document}